\documentclass[11pt,a4paper]{article}
\usepackage{cite}
\usepackage{graphicx}
\usepackage{epsf}%
\usepackage{epsfig}%
\usepackage{fancybox}%
\usepackage{pifont}%
\newlength{\capwidth}
\relax
\begin{document}
\begin{titlepage}
\flushright
L3 Internal Note 2184 \\
{\large{\today}}

\vspace*{10mm}

\begin{center}
{\bf\Large{Inclusion of Tau Anomalous Magnetic and Electric Dipole \\
Moments in the {\tt KORALZ} Monte Carlo}}

\vspace*{5mm}

{\large{T.~Paul$^1$ and Z.~W\c{a}s$^2$}}
\vglue 1cm
\begin{flushleft}
$^1$ Northeastern University, Boston, MA, USA.\\
$^2$ CERN, Theory Division, Geneva, Switzerland and Institute of Nuclear Physics, Krak\'{o}w, Poland\\
\end{flushleft}
\end{center}

\vspace*{30mm}

\begin{center}
{\bf{Abstract}} \\
We describe modifications made to {\tt KORALZ} version 4.03 in order to 
allow for anomalous magnetic and electric dipole moments of the $\tau$.
We discuss the verification of the method at LEP1 energies.
\end{center}

\parbox{0.8\textwidth}{%
}
\end{titlepage}
%
%
\section{Introduction} ~\label{sec:intro}
The analysis of radiative $\tau$ pair production provides a means to 
determine the anomalous magnetic and electric dipole moments of the 
$\tau$ at $q^2=0$.
An anomalous magnetic dipole moment at $q^2=0$, $F_2(0)$, or electric
dipole moment, $F_3(0)$, affects the total cross section for the process
$e^+e^-\rightarrow\tau^+\tau^-\gamma$ as well as the shape of energy and
angular distributions of the three final state particles~\cite{TTGTHEORY,BIEBEL96A,GRIFOLS91A}.
Previous experimental limits~\cite{GRIFOLS91A,LUSTERMAN,TAU96_L3} on $F_2(0)$ and 
$F_3(0)$ have been based on approximate calculations of the $e^+e^-\rightarrow\tau^+\tau^-\gamma$
cross section and photon energy distribution.  These calculations do not include the 
important effects of interference between anomalous and Standard Model amplitudes,
and furthermore can not be used to properly account for detector acceptance and selection cuts.

In order to address these problems, a tree level calculation of the squared matrix element
for the process $e^+e^-\rightarrow\tau^+\tau^-\gamma$ including the effects of non-zero 
$F_2(0)$ and $F_3(0)$  has been carried out. This matrix element calculation, 
which has been dubbed {\tt TTG}~\cite{TTGTHEORY}, may be used in the generation
of event samples with probabilistic weights, or may be applied in a Monte Carlo rejection 
method to produce events with weights of unity.  Perhaps more practically, the matrix element may 
be used to compute weights for any desired values of $F_2(0)$ or $F_3(0)$ given a set of 4-vectors 
for the final state particles in $e^+e^-\rightarrow\tau^+\tau^-\gamma$.  Thus it is straightforward
to interface this calculation with {\tt KORALZ}~\cite{KORALZ}, thereby providing all the 
Monte Carlo tools
necessary for a meaningful interpretation of the data.

This note describes the combination of {\tt KORALZ} with {\tt TTG}. First a brief 
description of the {\tt TTG} and {\tt KORALZ} programs is given. This is followed by 
description of how the two are interfaced and a description of 
the verification of the method for LEP1 energies.
Finally, we provide technical information on how to use the program.
%
%
\section{{ \tt TTG program }} ~\label{sec:ttg}
\begin{figure}[htbp!]
\begin{center}
\epsfig{file=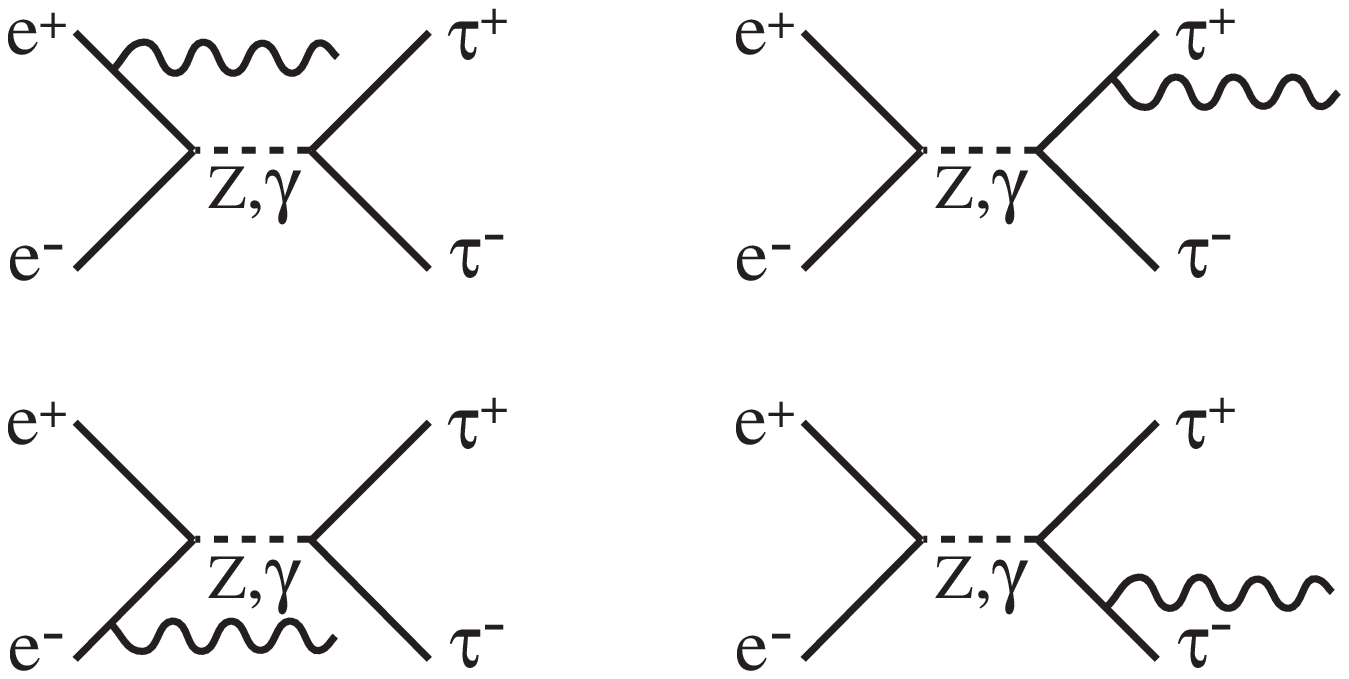,width=0.6\textwidth,clip=}
\vfill
\caption{\mbox{Diagrams contributing to $e^+ e^- \rightarrow \tau^+ \tau^-
                                         \gamma $}\label{fig:feynmann}}
\end{center}
\end{figure}

One may parametrize the effects of anomalous electromagnetic couplings in
$e^+e^-\rightarrow\tau^+\tau^-\gamma$  by 
replacing the usual $\gamma^\mu$ by more general Lorentz-invariant form of
the coupling of a tau to a photon\footnote{ in general there are 5 independent
lorentz invariant currents for spin 1/2 particle coupling to a photon.}:
\begin{equation}
\Gamma_\mu =          F_1 (q^2)           \gamma_\mu 
           + i \frac{ F_2 (q^2)}{2m_\tau} \sigma_{\mu\nu} q^\nu 
           -         F_3 (q^2)           \sigma_{\mu\nu} q^\nu \gamma_5,
\label{equ:formfactors}
\end{equation}
where $m_\tau$ is the mass of the $\tau$ lepton, and $q = p^\prime - p$ is
the momentum transfer.
As can be verified using the Gordon decomposition\cite{BJORKEN64A},
the $q^2$-dependent form-factors, $F_i(q^2)$, have familiar interpretations for $q^2=0$
and with the $\tau$ on mass-shell: 
$F_1(0) \equiv q_\tau$ is the electric charge of the tau,         
$F_2(0) \equiv a_\tau=(g-2)/2$ is the static anomalous magnetic moment of the tau                
(where $g$ is the gyromagnetic ratio), and       
$F_3(0) \equiv d_\tau/q_\tau$, where $d_\tau$ is the static electric dipole moment of the tau
and $q_\tau$ is its charge. 
Using this parametrization, we consider all the Standard Model and anomalous
amplitudes for the diagrams shown in Figure~\ref{fig:feynmann}.
The corresponding matrix element is then evaluated using the symbolic manipulation package FORM~\cite{FORM}
without making any simplifying assumptions.  In particular, no interference terms are
neglected and no fermion masses are assumed to be zero.  The matrix element
is available from the authors in the form of a FORTRAN subroutine.

Following an initial presentation of these results~\cite{SWAIN96A},
an analytical calculation for the process
$e^+e^-\rightarrow\tau^+\tau^-\gamma$ was carried out~\cite{BIEBEL96A}.
This calculation makes some approximations, but importantly 
does not assume zero tau mass and does not neglect interference between
Standard Model and anomalous final states.  This provides with a means
to crosscheck the results of {\tt TTG}.
The {\tt TTG} program and the crosschecks are described in detail
in reference~\cite{TTGTHEORY}.
%
%
\section{{\tt KORALZ} Program} ~\label{sec:koralz}

In this section we briefly review the properties of {\tt KORALZ} which are relevant to the
modifications we describe in section~\ref{sec:merging}.
The algorithm employed by {\tt KORALZ} for generation of $e^+e^- \to \tau^+ \tau^- (\gamma)$,
including radiative corrections and $\tau$ decay, is described in detail in 
reference \cite{KORALZ}. 

Real photon radiation in {\tt KORALZ} is controlled by the {\tt KEYRAD} flag.
The program may be run at Born level ({\tt KEYRAD=0}),
may include order $\alpha$ QED corrections ({\tt KEYRAD=1}), or  
order $\alpha^2$ QED corrections including exclusive exponentiation
({\tt KEYRAD=12}).  The Born level differential distribution is used as
a starting point in calculating the matrix
element at ${\cal O}(\alpha)$ and ${\cal O}(\alpha^2$).

In the case of {\tt KEYRAD=12}, the user may switch on and off 
contributions from initial state radiation (ISR) and final state
radiation (FSR) using the {\tt NPAR(12)} card.  
For testing purposes, three additional {\tt KEYRAD} options have 
also been introduced which allow one to turn on (or off) ISR, FSR, 
and interference for the case of single bremsstrahlung, as summarized below.\\
\vspace{2mm}
\begin{center}
\begin{tabular}{|ll|} \hline
{\tt KEYRAD=0 }                      & Born level                                     \\
{\tt KEYRAD=1 }                      & ${\cal O}(\alpha)$, ISR, FSR, interference     \\
{\tt KEYRAD=2 }                      & ${\cal O}(\alpha)$, ISR, FSR                   \\
{\tt KEYRAD=3 }                      & ${\cal O}(\alpha)$, ISR                        \\
{\tt KEYRAD=4 }                      & ${\cal O}(\alpha)$, FSR                        \\
{\tt KEYRAD=12, NPAR(12)=1000011}    & ${\cal O}(\alpha^2)+$~exponentiation, ISR, FSR \\
{\tt KEYRAD=12, NPAR(12)=1000001}    & ${\cal O}(\alpha^2)+$~exponentiation, ISR      \\
{\tt KEYRAD=12, NPAR(12)=1000010}    & ${\cal O}(\alpha^2)+$~exponentiation, FSR      \\ \hline
\end{tabular}
\end{center}
%

%
%
\section{Merging {\tt TTG} with {\tt KORALZ}} ~\label{sec:merging}
{\tt KORALZ} and {\tt TTG} have been merged such that for each event generated by 
{\tt KORALZ}, a weight is computed by {\tt TTG} for a given $F_2(0)$ or $F_3(0)$.
Details of the weight calculation and how the information may be 
accessed is given in section~\ref{sec:howto}.  Since we are only 
interested in events with photons, events without photons can be rejected
by setting the {\tt KORALZ} internal weight to zero \footnote{For that purpose, 
the internal input parameter {\tt IRECSOFT} in routine {\tt kzphynew(XPAR,NPAR)} 
should be set to 1.}.  In this case, 
the total cross section given at the end of the {\tt KORALZ} run will include
only contributions from configurations with a real hard photon above
the {\tt KORALZ} internal parameters {\tt xk0} or {\tt vvmin}~\footnote{Cross 
sections corresponding to realistic cuts on minimal photon energies 
will not be affected by the choice of these parameters.}.

Since {\tt TTG} provides an ${\cal O}(\alpha)$ calculation,
this procedure is straightforward and unambiguous for {\tt KEYRAD = 1,2,3,4},
where there is at most a single bremsstrahlung photon in the event.
In this case, the {\tt KORALZ}/{\tt TTG} program simply works as a single 
bremsstrahlung generator with anomalous contributions included.

For multiple photon events ({\tt KEYRAD=12}), the situation is not as
simple because the weight factor (see equation~\ref{eq:weight} of section
~\ref{sec:multi}) for the anomalous contribution can no longer be
calculated in a direct way. In this case we need to
rely on a reduction procedure in which
all photons except for the one with largest $p_T$ are incorporated
into the 4-momenta of effective initial or final state leptons.
This approach is founded on the basic factorization properties of  
QED. In the infrared limit, as well as for important regions of phase space
which give leading log corrections, the matrix element
can be written (up to non-leading terms) as a product
of the Born level matrix element multiplied by factors $S_i$ corresponding
to photon(s) emission. The $S_i$ are independent
from the particular hard process under consideration. A similar property 
holds for phase space. 
See for example reference \cite{zakopane-school} for an introductory presentation
and references. Here we will assume that anomalous couplings
of the photon to the $\tau$ do not affect these properties and that their effect
can be described as correction or perturbation to the individual factors  
$S_i$~~\footnote{In fact, as it will be explained later, we will take 
this perturbation for only the photon of the highest
$p_T$ with respect to leptons. We will also assume that anomalous contributions
are not of the infrared divergent or collinear divergent type.}.
Thus if one assumes that non-leading corrections are small and can be neglected,
then such a reduction procedure may be combined with the calculations of {\tt TTG} in
order to account for anomalous contributions in the case of multiple photon radiation.
%

%

We now present details of the reduction procedure just discussed.
This procedure is performed in the routine {\tt WTANOM}, which 
is called in the case of flag {\tt IFKALIN=2}.  For each 
generated event which contains more than one real photon,
the following algorithm is applied:

\begin{itemize}

\item The invariant mass, $m_i^k$, for each pair of particles containing 
a photon, $i=1...n$, and a lepton, $k=1...4$ where 
$1,2 = e^\pm$ and $3,4=\tau^\pm$, is calculated.  

\item The masses $m_i^k$ are multiplied by the square of the sum of the 
photon and $Z$ propagators. The energy transfer for $k=3,4$ is taken to be the center-of-mass
energy, whereas for $k=1,2$ this transfer is reduced by $1-E_i/E_{\mathrm{beam}}$.  
This increases the value of $m_i^k$ for the case of a hard photon paired with 
a beam electron, reflecting the fact that a narrow resonance cuts off contributions
from hard ISR.  


\item The minimum $m_i$ out of $m_i^{k=1..4}$ is selected.

\item The maximum $m$ of the $m_i$ is selected.
The corresponding photon is stored as the highest $p_T$ photon;
this is the photon that will be passed to {\tt TTG} to compute 
the weights corresponding to anomalous moments.
The four-momentum of each remaining photon, $i$, is added to one of the 
final state $\tau$'s or subtracted from one of the initial state
$e$'s, depending on which $m_i^{k=1..4}$ is the smallest.  The resulting
lepton-photon combinations are referred to as ``effective'' $\tau$'s or 
beams.

\item The 4-momenta of the effective $\tau$'s and beams as
well as the highest $p_T$ photon are boosted into the rest frame 
of the effective beams.  We call this the rest frame of the effective reaction.

%
\item The 4-momenta of the boosted effective beams are modified such that 
they are back-to-back in the rest frame of the effective reaction, and 
are consistent with the electron mass.


\item The 4-momenta of the effective $\tau$'s are boosted into the 
rest frame of the $\tau$ pair. 

%
\item The 4-momenta of these boosted effective $\tau$'s are modified
such that they are back-to-back and consistent with the $\tau$ mass.

%
\item These modified $\tau$ 4-vectors are boosted back into the frame 
of the effective reaction.


\end{itemize}
At this point we have constructed the kinematical configuration of
the reaction $e^+e^- \to \tau^+ \tau^- \gamma$, ensuring all the leptons 
are on mass shell.  These 4-vectors may then be used by {\tt TTG} to calculate the 
standard model or anomalous matrix element for this process in an 
unambiguous way.

In the final step of the event generation, the $\tau$ decay is 
simulated using TAUOLA~\cite{TAUOLA}.  In the calculation of 
spin effects which occurs at this point, any effects of 
anomalous contributions are neglected.

A rigorous evaluation of the quality of the algorithm described 
above would require careful comparison of the generated distributions
with those of the exact matrix element calculations including 
anomalous contributions to at least ${\cal O}(\alpha ^2)$.
This is impossible at the moment, but similar tests for the process
$e+e- \to \nu \bar \nu \gamma$'s were performed in \cite{COLAS} and 
the approximation worked quite well.
  

The above procedure introduces systematic uncertainties only in 
the {\it anomalous contributions} to the distribution of the 
final state particles (except at the single photon level, where no 
such systematic is introduced).  Moreover, these 
uncertainties do not affect corrections at the leading-log QED 
level or infrared/collinear regions of the distribution of 
radiated photons.  Only the next-to-leading-log contributions 
of higher orders are affected.  

%
%
%
\begin{figure}[htbp!]
\begin{center}
\epsfig{file=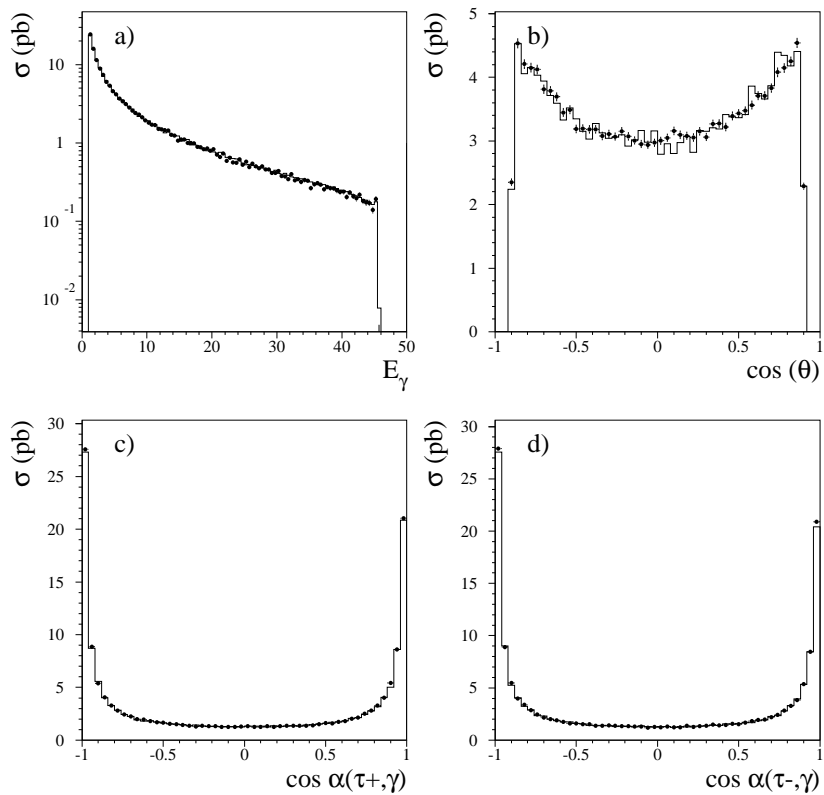,width=0.5\textheight,clip=}
\vfill
\caption{Comparison of predictions of {\tt TTG} (histogram) with {\tt KORALZ} (dots) for
a) photon energy, b) angle between photon and beam electron, c) angle between photon 
and $\tau^+$, d) angle between photon and $\tau^-$. The cuts defined 
in section~\ref{sec:crosschecks} have been applied.
\label{fig:kz_ttg}}
\end{center}
\end{figure}
\begin{figure}[htbp!]
\begin{center}
\epsfig{file=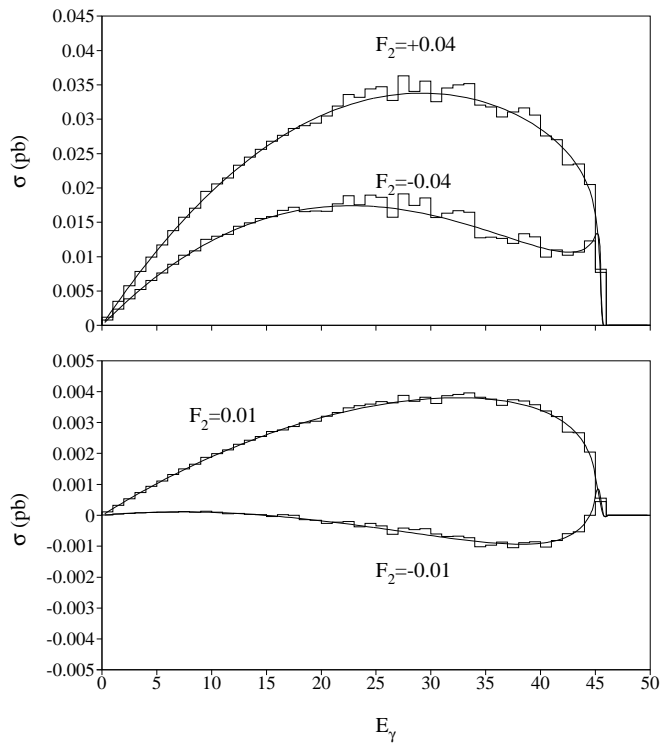,width=0.4\textheight,clip=}
\vfill
\caption{The anomalous contribution to the cross section as a function of 
energy for several values of $F_2(0)$.  The histogram is the result of 
the {\tt {\tt KORALZ}/TTG} Monte Carlo and the curve is from the analytical calculation.
No cuts have been applied.
\label{fig:x_mdm}}
\end{center}
\end{figure}
\section{Crosschecks at ${\cal{O}(\alpha)}$} ~\label{sec:crosschecks}
As a technical crosscheck, we compare the predictions of {\tt TTG} with settings
$F_2(0) = F_3(0) = 0$ with those of {\tt KORALZ}.  Since {\tt TTG} provides only 
an ${\cal O}(\alpha)$ calculation, {\tt KORALZ} is run with {\tt KEYRAD=1} so that only
single photon radiation is considered.  In order to prevent infrared 
divergences in the {\tt TTG} calculation, we impose cuts on the minimum photon 
energy ($E_{\mathrm{min}}>$~1~GeV), the angle between the photon and the beam
electron ($\left | \cos\theta \right | <$~0.9), and the angle between the photon and closest 
tau ($\cos\alpha_{\mathrm{min}}<$~0.995).  With these cuts, the total cross
sections predicted by {\tt KORALZ} and {\tt TTG} agree to with about $0.1\%$.  
Figure~\ref{fig:kz_ttg} shows a comparison of the energy and angular distributions
computed by the two programs.

Next we verify that the combined {\tt KORALZ}/{\tt TTG} program correctly calculates the 
anomalous contribution to the cross section for $F_2(0) \neq 0$ or 
$F_3(0) \neq 0$.  For this check, we make use of the analytical calculation
described in reference~\cite{BIEBEL96A}.  This calculation neglects anomalous
contributions from initial-final state interference, from $\gamma Z$ interference,
and from $\gamma$ exchange, so from purposes of comparison we remove these 
terms from the {\tt TTG} calculation and we run {\tt KORALZ} with {\tt KEYRAD=4}, in which case
only FSR is considered.  To remove any ambiguity concerning the validity of 
comparing the non-QED genuine weak corrections computed by {\tt KORALZ} with the improved Born 
approximation approach used in the analytical calculation, we set {\tt KEYGSW=1} in 
{\tt KORALZ} and use the Born approximation.  The anomalous contribution to the 
cross section computed by {\tt KORALZ}/{\tt TTG} agrees with that of the analytical 
calculation to $1\%$.  Figure~\ref{fig:x_mdm} shows, for several 
values of $F_2(0)$, a comparison of the 
anomalous contribution to the photon energy spectrum computed by {\tt KORALZ}/{\tt TTG} 
with the predictions of the analytical calculation.
%

%
%
\section{Results for Multiple Photon Radiation}~\label{sec:multi}
We now present numerical results of the {\tt KORALZ}/{\tt TTG} program with 
multiple photon radiation included.  The goal is to demonstrate that 
the reduction scheme described in section~\ref{sec:merging} gives 
results which are consistent with expectations from the ${\cal O}(\alpha)$
calculation, and to indicate how one might estimate the systematic
errors associated {\tt KORALZ}/{\tt TTG} simulation including multiple photon radiation.

\begin{figure}[htbp!]
\begin{center}
\epsfig{file=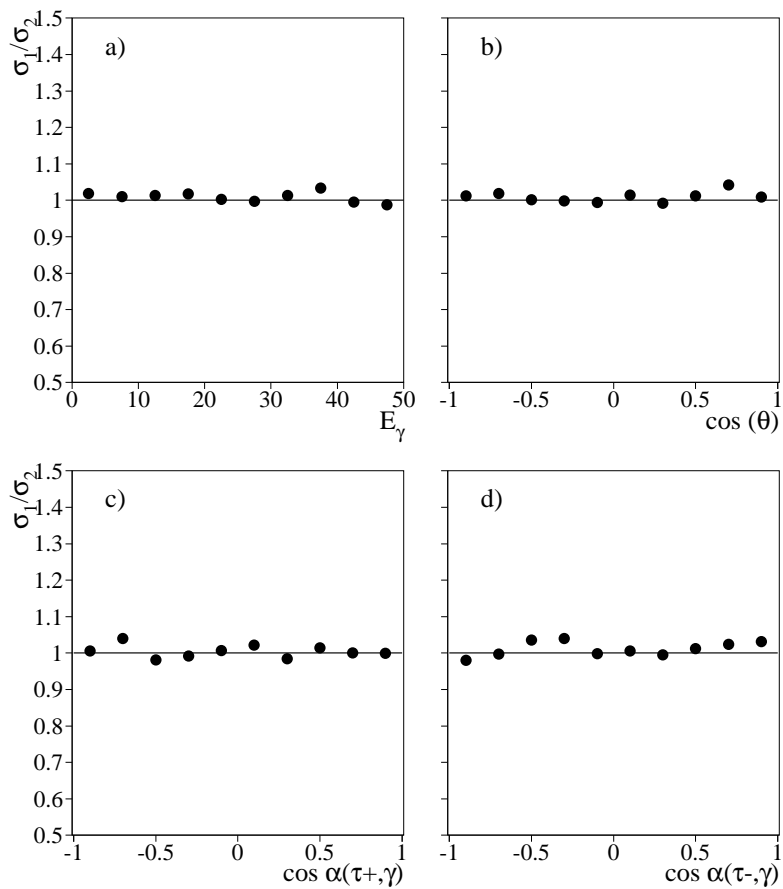,width=0.5\textheight,clip=}
\vfill
\caption{Ratios of differential cross sections predicted using 
{\tt KEYRAD=1} (ISR,FSR,interference included) to that predicted
using {\tt KEYRAD=2} (ISR,FSR) for a) photon energy, b) angle between
photon and beam electron c)  angle between photon 
and $\tau^+$, d) angle between photon and $\tau^-$.
The value $F_2(0)=+0.04$ has been used and
the cuts specified in section~\ref{sec:crosschecks} have been applied.
\label{fig:interf}}
\end{center}
\end{figure}

First, we check that the interference between initial and final 
state bremsstrahlung at ${\cal O}(\alpha)$ contributes negligibly to our 
observables.  To this end, we compare our observables as they are 
computed by {\tt KORALZ}/{\tt TTG} using {\tt KEYRAD=1} with the results using
{\tt KEYRAD=2}.  Figure~\ref{fig:interf} shows the ratios for the two
calculations.
It is necessary to check the effects of interference at ${\cal O}(\alpha)$,
as interference is not included in the simulation of multiple photon
radiation ({\tt KEYRAD=12}).  As the contribution from interference turns
out to be very small for the single photon calculation, we may safely
proceed with our {\tt KEYRAD=12} checks with
further consideration of possible ISR/FSR interference effects.

From now on we will exploit
the fact that, to a good approximation, the single (or highest $p_T$) photon
distribution can be represented as a simple sum of the ISR and FSR contributions.
For each event, the weight is calculated by {\tt TTG} to be
\begin{equation} \label{eq:weight}
w = {|M_I+M_F+M_A|^2\over |M_I+M_F|^2}
\end{equation}
Here $M_I$, $M_F$, and $M_A$ denote respectively the matrix element for photon
emission from initial, final states and from the final state tau through
anomalous coupling.  Let us denote the distributions which include
anomalous contributions as:
\begin{eqnarray}
d\sigma_I^A=d\sigma_I\; w  \nonumber\\
d\sigma_F^A=d\sigma_F\; w  \nonumber\\ 
d\sigma^A=d\sigma\; w      \nonumber
\end{eqnarray}
with
\begin{eqnarray}
d\sigma_I=\left | {\cal M}_I \right | ^2 d\Omega \nonumber  \\ 
d\sigma_F=\left | {\cal M}_F \right | ^2 d\Omega \nonumber
\end{eqnarray}
where ${\cal M}$ is the matrix element and $d\Omega$ is the 
invariant phase space element.
Thanks to the smallness of the ISR/FSR interference, we can 
say to good approximation that
\begin{equation}
d\sigma=d\sigma_I+d\sigma_F
\end{equation}
and as a consequence
\begin{equation}
d\sigma^A=d\sigma_I^A+d\sigma_F^A.
\end{equation}

In this sense we can separate total anomalous contributions into
independent contribution from initial and final states. 

\begin{figure}[htbp!]
\begin{center}
\epsfig{file=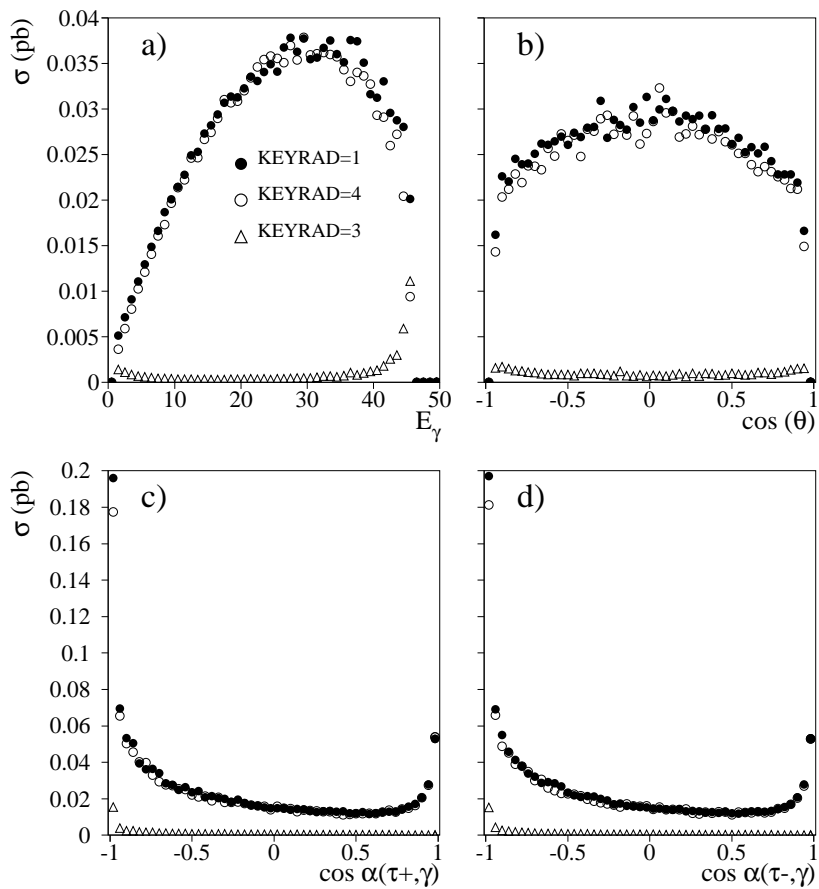,width=0.5\textheight,clip=}
\vfill
\caption{Anomalous contribution to differential cross section for the case of 
ISR ({\tt KEYRAD=3}), FSR ({\tt KEYRAD=4}), and ISR+FSR+interference
({\tt KEYRAD=1}) shown as a function of a) photon energy, b) angle between
photon and beam electron c)  angle between photon 
and $\tau^+$, d) angle between photon and $\tau^-$. The cuts
specified in section~\ref{sec:crosschecks} have been applied,
and the value $F_2(0)=+0.04$ has been used.
\label{fig:key_1_3_4}}
\end{center}
\end{figure}

In Figure~\ref{fig:key_1_3_4}, we compare the single bremsstrahlung calculations
of the {\it anomalous} contribution to the differential cross section
including ISR only ({\tt KEYRAD=3}), FSR only ({\tt KEYRAD=4}), and all 
contributions ({\tt KEYRAD=1}).  Note that the bulk of the anomalous
contribution arises from final state radiation; this is expected, since
the photons from  phase space regions where ISR dominates produce rather
small anomalous corrections to the amplitudes.

\begin{figure}[htbp!]
\begin{center}
\epsfig{file=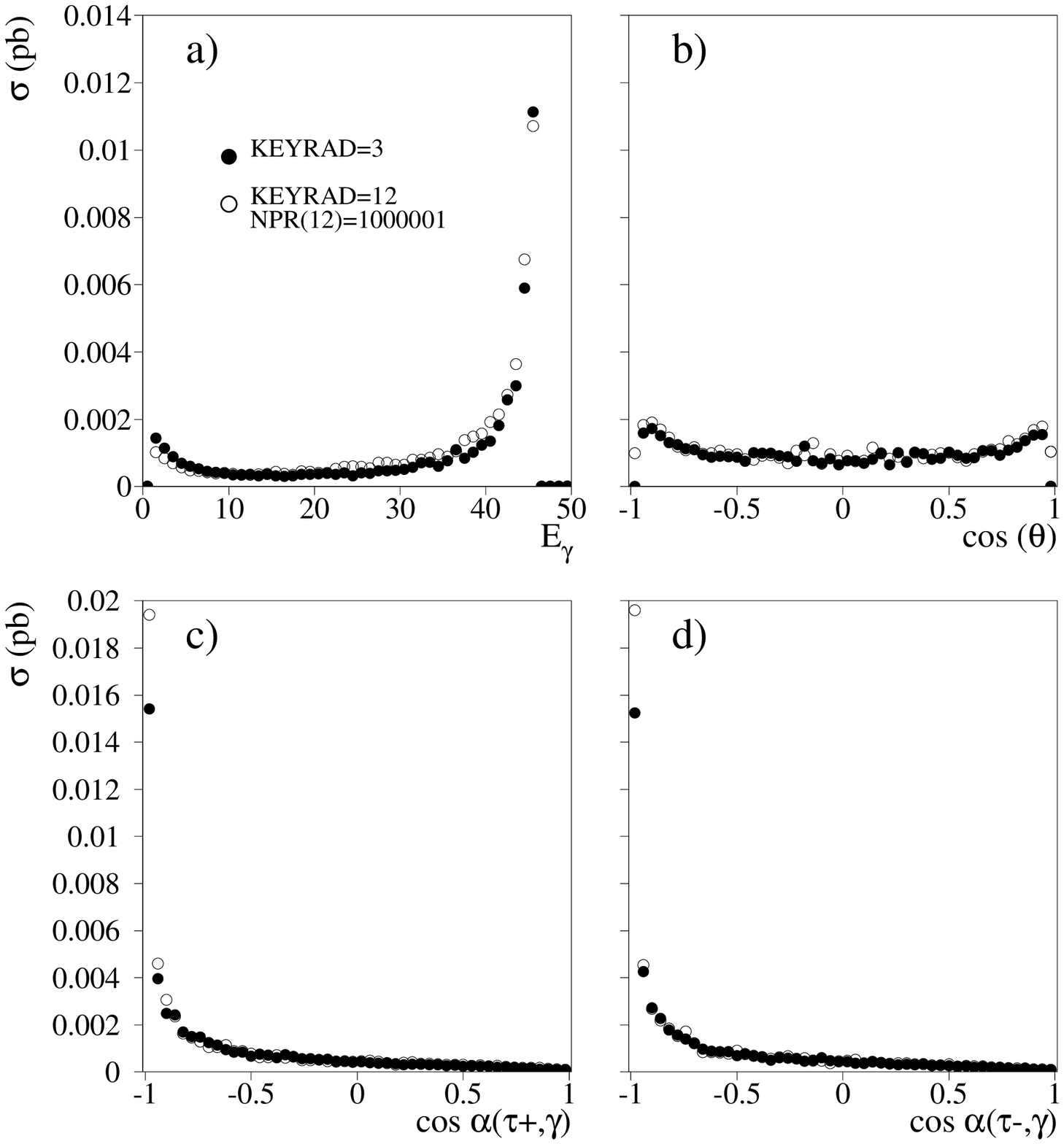,width=0.5\textheight,clip=}
\vfill
\caption{Comparison of the anomalous contribution to the differential
cross section for single and multiple bremsstrahlung including only
ISR ({\tt KEYRAD=3} compared to {KEYRAD=12,NPR(12)=1000001}).  
Differential cross sections are shown as a function of
a) photon energy, b) angle between
photon and beam electron c)  angle between photon 
and $\tau^+$, d) angle between photon and $\tau^-$. The cuts
specified in section~\ref{sec:crosschecks} have been applied,
and the value $F_2(0)=+0.04$ has been used.
\label{fig:key_3_12}}
\end{center}
\end{figure}

\begin{figure}[htbp!]
\begin{center}
\epsfig{file=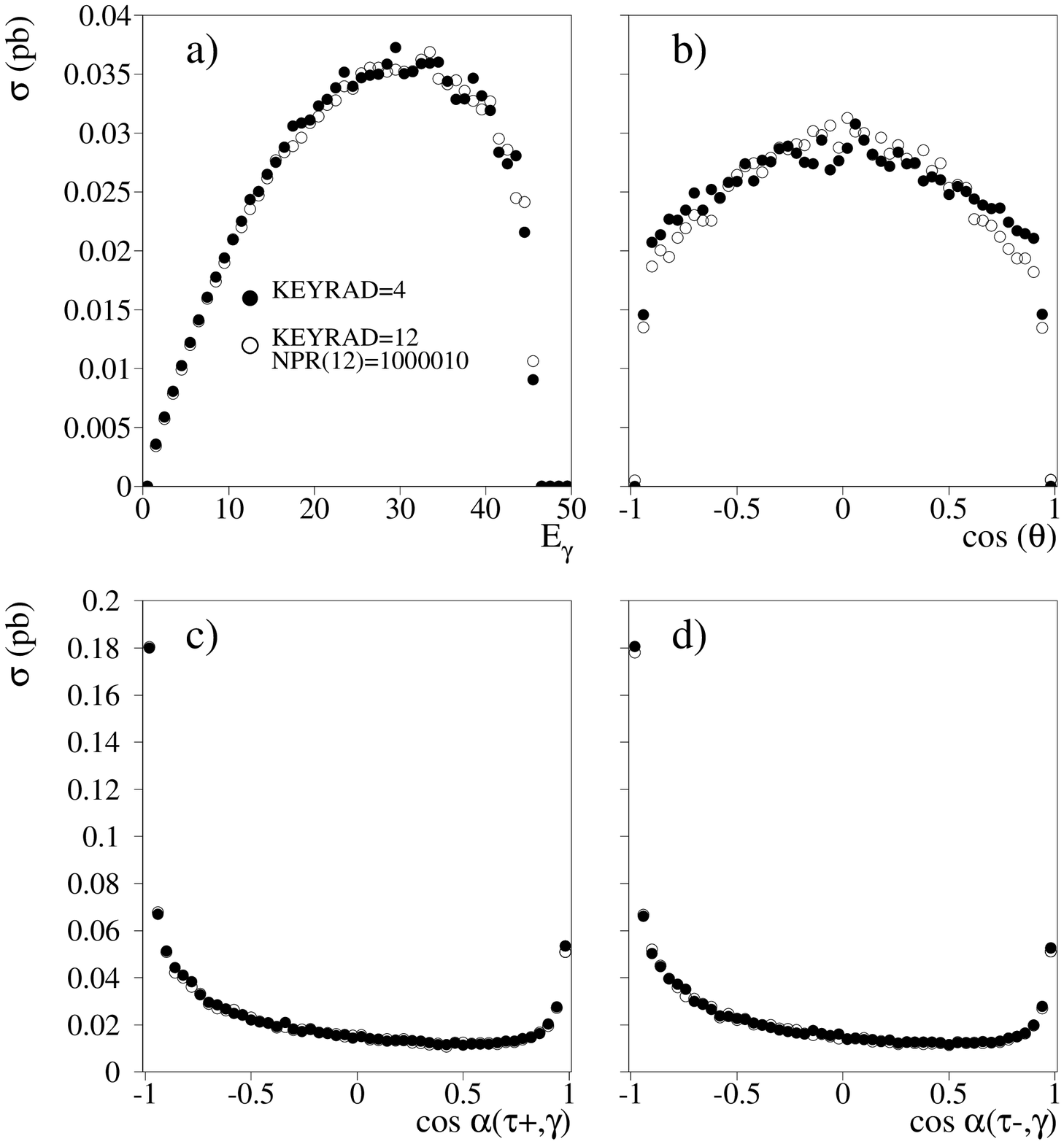,width=0.5\textheight,clip=}
\vfill
\caption{Comparison of the anomalous contribution to the differential
cross section for single and multiple bremsstrahlung including only
FSR ({\tt KEYRAD=4} compared to {KEYRAD=12,NPR(12)=1000010}).  
Differential cross sections are shown as a function of
a) photon energy, b) angle between
photon and beam electron c)  angle between photon 
and $\tau^+$, d) angle between photon and $\tau^-$. The cuts
specified in section~\ref{sec:crosschecks} have been applied,
and the value $F_2(0)=+0.04$ has been used.
\label{fig:key_4_12}}
\end{center}
\end{figure}

\begin{figure}[htbp!]
\begin{center}
\epsfig{file=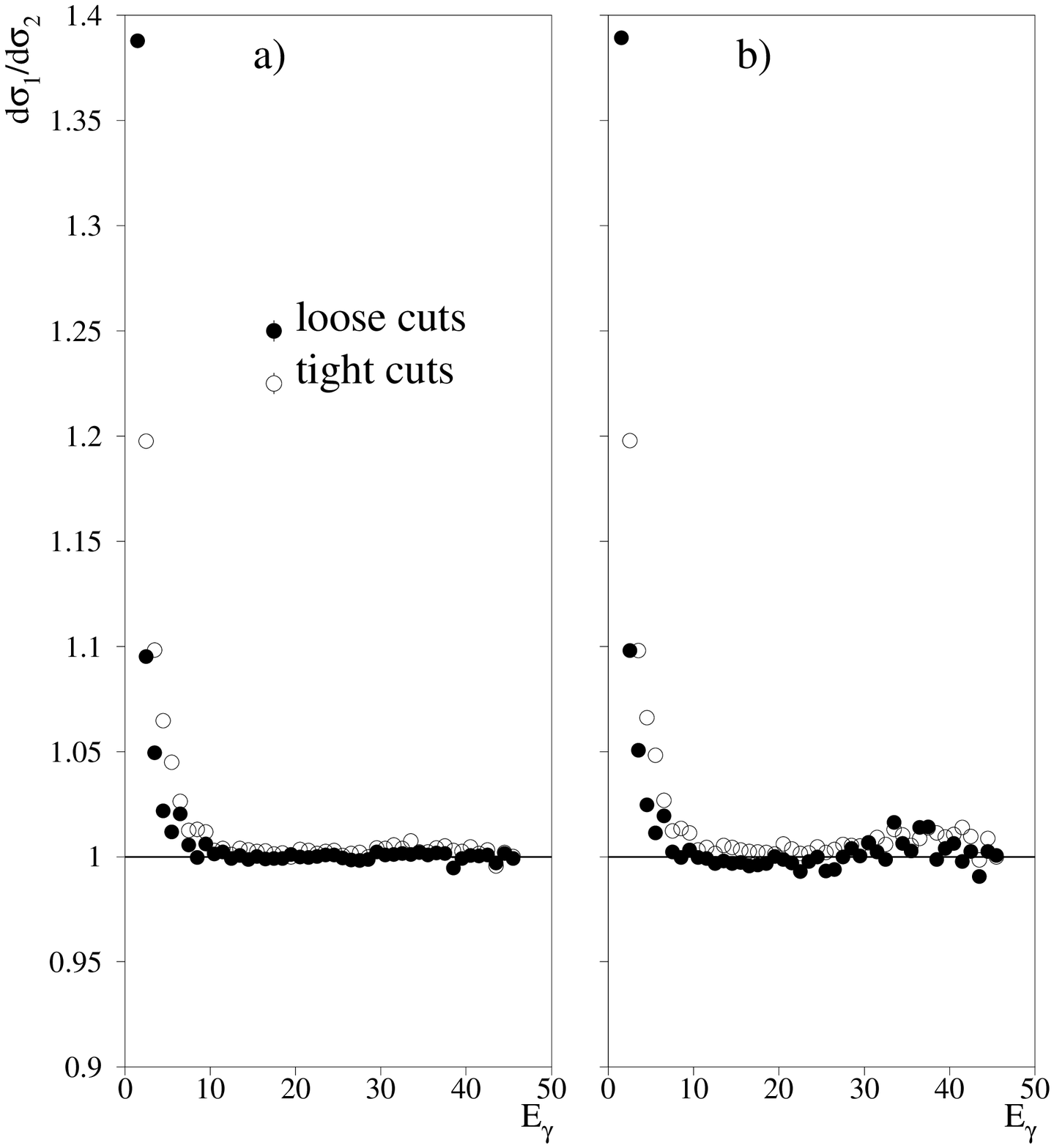,width=0.6\textwidth,clip=}
\vfill
\caption{a) Ratio of the anomalous contribution to the differential
cross section as a function of photon energy as computed using the
reduction scheme to that computed not using it.  Exactly one photon is 
required to satisfy $E_\gamma>1$GeV, $| cos(\theta) | <0.9$, 
$\cos(\alpha_{\tau^+,\gamma})<0.995$, and $\cos(\alpha_{\tau^-,\gamma})<0.995$
(loose cuts), or  $E_\gamma>2.5$GeV, $| cos(\theta) | <0.7$, 
$\cos(\alpha_{\tau^+,\gamma})<0.95$, and $\cos(\alpha_{\tau^-,\gamma})<0.95$
(tight cuts).  b) The same ratio for the case of one or more photons satisfying
the loose or tight cuts.  Only the highest energy photon is used to compute
the weight in the case of no reduction.  For loose cuts, $3.9\%$ of {\it selected}
events have more than one photon, and for tight cuts, $1.5\%$ have more than one.
The value $F_2(0)=+0.04$ has been used.
\label{fig:systematic}}
\end{center}
\end{figure}

Next, we compare the anomalous contribution to the cross section calculated
assuming single brems\-strah\-lung with that including multiple bremsstrahlung.
Figure~\ref{fig:key_3_12} shows this comparison for the case of 
ISR alone, and Figure~\ref{fig:key_4_12} shows the same comparison for the
case of FSR.  In the case of multiple bremsstrahlung, the reduction
procedure of section~\ref{sec:merging} has been employed, and the 
energies and angles plotted are those of effective reaction \footnote{As we
will show in Figure~\ref{fig:systematic}, it makes essentially no difference
whether we use the angles for the effective reaction or the angles
in the real particles in the laboratory system.}.
We can see that the difference between the two calculations is not
dramatic~\footnote{Keep in mind that the the {\it total} anomalous cross section
shown in these plots (corresponding to $F_2=0.04$) produces only a 1\% or so 
effect on the total $\tau \tau \gamma$ cross section.}, especially for the 
more sensitive regions of the distributions.

\begin{figure}[htbp!]
\begin{center}
\epsfig{file=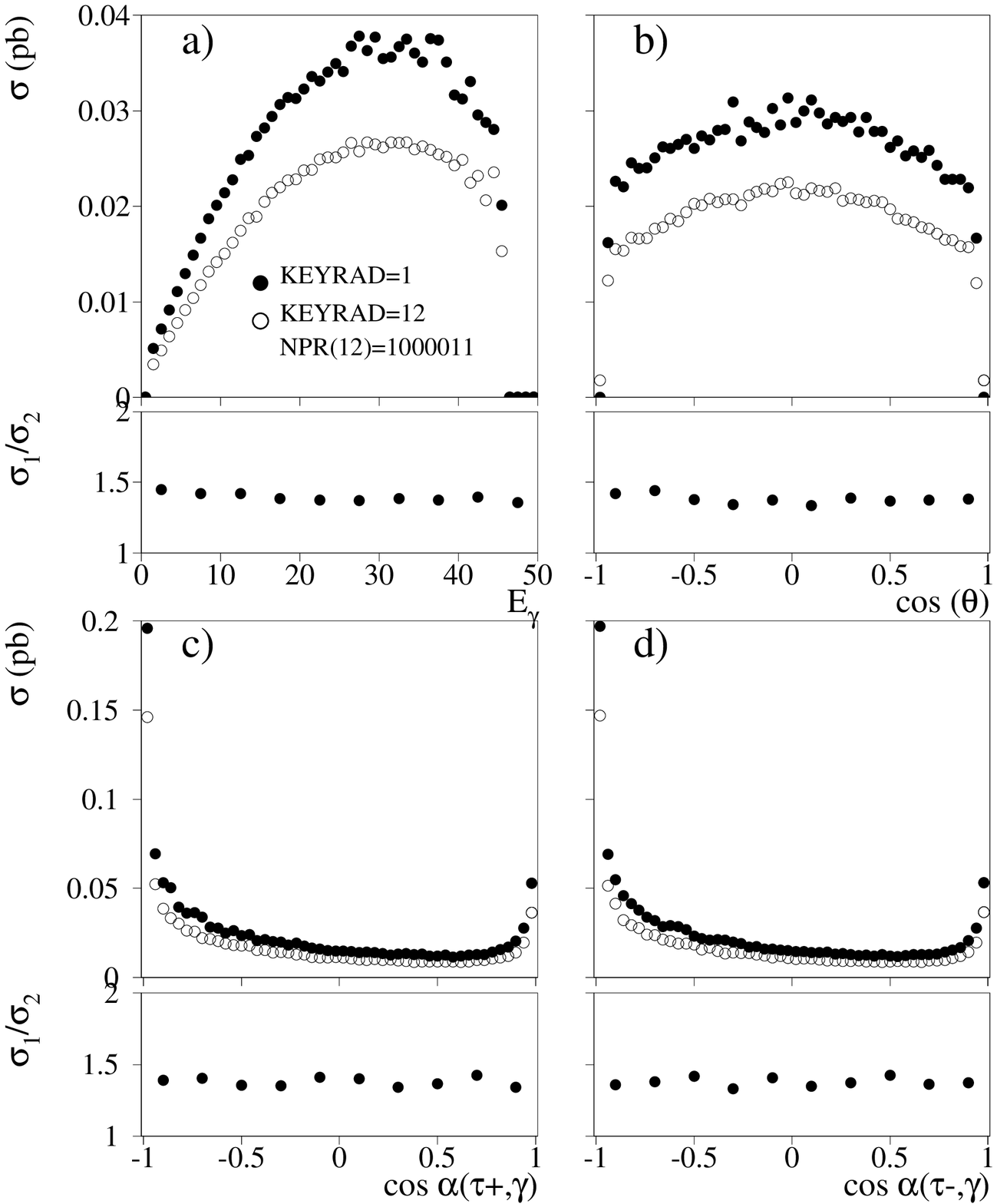,width=0.98\textwidth,clip=}
\vfill
\caption{Comparison of the anomalous contribution to the differential
cross section for single and multiple bremsstrahlung including 
both ISR and FSR ({\tt KEYRAD=1} compared to {KEYRAD=12,NPR(12)=1000011}).  
The larger plots show these two distributions, while the lower plots 
show the ratio of the two.  
Differential cross sections are shown as a function of
a) photon energy, b) angle between
photon and beam electron c)  angle between photon 
and $\tau^+$, d) angle between photon and $\tau^-$. The cuts
specified in section~\ref{sec:crosschecks} have been applied,
and the value $F_2(0)=+0.04$ has been used.
\label{fig:key_12_12}}
\end{center}
\end{figure}

Finally, Figure~\ref{fig:key_12_12}
gives a comparison of the single and multiple bremsstrahlung calculations
for the anomalous cross section
including both ISR and FSR (and interference, in the single 
bremsstrahlung case).  Again, the reduction scheme has been 
applied in the case of multiple bremsstrahlung.  As expected,
the overall anomalous cross section is suppressed by including
multiple photon radiation in the initial state, but the overall
shape is not strongly affected.
%
%
%
%
%

From these consistency checks, we conclude that the reduction 
algorithm gives sensible results.  Aside from the expected
overall scaling of the cross section due to ISR, 
the effects of higher order
corrections on anomalous contributions to the differential 
cross section are small compared to the anomalous
contributions themselves.  This suggests that the related
systematic errors on measurement of $F_2(0)$ and $F_3(0)$ should
be small at LEP1 energies.
%

We may also estimate the size of systematic errors associated
with higher order corrections to anomalous contributions
by simulating events including
multiple bremsstrahlung and comparing the anomalous distributions
obtained using the reduction procedure to those obtained 
without using it.  In order to do this, we select only events
with {\it exactly} one photon satisfying certain energy and angle 
requirements.  Two weights are then computed, the first using the
4-vectors of the selected photon and the two taus, and the second
using the reduced 4-vectors which take into account any additional
soft photons that may be present (but which do not pass the 
selection). The ratio of the anomalous photon energy distributions for
these two approaches is shown in 
Figure~\ref{fig:systematic}a for two sets of selection cuts.  
In all cases, the photon energy stored in the histogram is that
of the selected photon, not the photon energy seen in the effective
frame.  The discrepancy is significant at low energies, but 
essentially vanishes in the interesting high energy regions.
Figure~\ref{fig:systematic}b is a similar comparison for a slightly
different selection; in this case, we ask for one {\it or more} photons
to pass the selection criteria, and compute one weight using only 
the highest energy selected photon and the taus and and the second 
weight using the reduction method.  Again, there is only a small 
difference between the two approaches in the high energy regions.

Despite the fact that the effects of multiple bremsstrahlung appear small
in the regions of interest, they are nonetheless included in the 
{\tt KORALZ}/{\tt TTG} Monte Carlo.  This simplifies the selection and fitting.
Finally, let us stress that our estimation of the systematic error
is valid only for the observables, cuts, and center-of-mass energies defined
here.  For other choices, checks similar to those presented here should be performed.

%
 

y%
%
\section{How to Use the Program} ~\label{sec:howto}
As discussed in sections~\ref{sec:intro} and~\ref{sec:merging}, the strategy to
account for anomalous magnetic and electric dipole moments involves using 
{\tt KORALZ} to generate $\tau$ pairs with one or more radiated photons, applying
a reduction procedure in the case of multiple photon radiation, and 
computing a weight, $w$,  {\it for the event} as,
\begin{equation}
w =  \frac{| {\cal M}_\mathrm{ano} |^2} {| {\cal M}_\mathrm{SM} | ^2},
\end{equation}
where ${\cal M}_\mathrm{ano}$ is the matrix element computed by {\tt TTG} 
for $F_2(0) \neq 0$ and/or $F_3(0) \neq 0$ using the 4-vectors for the 
taus and the photon, and 
${\cal M}_\mathrm{SM}$ is the matrix element, also computed by {\tt TTG},
for the case of $F_2(0)=F_3(0)=0$.

The calculation of these weights is activated by 
setting the card {\tt IFKALIN=2}. This is transmitted from 
the main program via the {\tt KORALZ} input parameter {\tt NPAR(15)}.  
If this card is set, then {\tt KORALZ} initializes {\tt TTG} by calling the
routine {\tt ANOMINI\_L3}.  Constants of nature are passed from {\tt KORALZ} to 
{\tt ANOMINI\_L3} with the help of the routine {\tt KZ\_STOREPARMS}.
The reduction procedure described in section~\ref{sec:merging} is performed
for each event in the routine {\tt WTANOM}.  After reduction, the actual weights 
for anomalous couplings are calculated by calling the routine
{\tt FU\_L3}.  All devices necessary for
the importance sampling algorithm which minimizes the statistical divergence
on difference in distributions with and without anomalous couplings
are in place.  

Additional options for {\tt TTG} are anticipated in the common block
{\tt TTG\_USER}. Such options are currently set in the
routine {\tt kzphynew(XPAR,NPAR) },
but there are no connections (yet) to the {\tt KORALZ} matrix input parameters
{\tt XPAR, NPAR}, though it is straightforward to implement this.
For the moment, one may set the following flags in
the {\tt TTG\_USER} common:\\
\begin{center}
\begin{tabular}{|ll|} \hline
{\tt IF1}             & $=1$ to compute weights for $F_2(0)$  \\
{\tt IF2}             & $=1$ to compute weights for $F_3(0)$  \\
{\tt ISFL}            & {\tt TTG} ``simple'' flag \\ \hline
\end{tabular}
\end{center}
\vspace*{2mm}
where one or of both {\tt IF1} and {\tt IF2} may be set, and where
{\tt ISFL} may have the following settings: \\
\vspace*{2mm}
\begin{center}
\begin{tabular}{|ll|} \hline
{\tt ISFL}~$=-1$                & {\tt TTG} computes
                                  {\em only} terms with anomalous contributions \\
{\tt ISFL}~$=0$                 & {\tt TTG} includes all terms \\
{\tt ISFL}~$=1$                 & {\tt TTG} uses the 
                                   approximation of reference~\cite{BIEBEL96A}  \\ \hline
\end{tabular}
\end{center}
\vspace*{2mm}

In order to provide the user with enough information to retrieve $w$ 
for a given event for any $F_2(0)$ or $F_3(0)$, we take advantage of 
the fact that, for each event, we may write $w$ as a quadratic function of the 
anomalous couplings:
\begin{equation}
w = \alpha F_2^2(0) + \beta F_2(0) + \gamma F_3^2(0) + \delta F_3(0) + \epsilon.
\end{equation}
When {\tt FU\_L3} is called, 
these 5 constants are stored in the common block\footnote{This is similar to
the way anomalous $\nu \bar \nu \gamma$ information is stored~\cite{NUNUPREP}.}
{\tt common /kalinout/ wtkal(6)}, with the following assignments:
\vspace*{2mm}

\begin{center}
\begin{tabular}{|ll|} \hline
{\tt wtkal(1)}  &  not used here (see \cite{NUNUPREP}) \\ 
{\tt wtkal(2)}  &  $\epsilon$\\
{\tt wtakl(3)}  &  $\alpha$  \\    
{\tt wtkal(4)}  &  $\beta$   \\
{\tt wtkal(5)}  &  $\gamma$  \\
{\tt wtkal(6)}  &  $\delta$  \\
\hline
\end{tabular}
\end{center}
\vspace*{2mm}

The user is then free to calculate $w$ for whatever combination of $F_2(0)$ 
and $F_3(0)$ is desired.
Note that in practice we set $\epsilon=1$, since anomalous terms must vanish
for $F_2(0)=F_3(0)=0$, and $\delta=0$, as the interference between 
standard model and anomalous amplitudes vanishes in the case of 
radiation from an electric dipole moment.  These shortcuts save 
substantial CPU time.
%

%
%
\section*{Acknowledgements}
ZW would like to thank L3 group of ETH Zurich for support
while this work was performed.  Support was also provided by
Polish Government grant KBN 2P03B17210.  TP would like to
acknowledge the support of the National Science Foundation.
\clearpage
%


\end{document}